\documentclass[sigconf]{acmart}

\usepackage{times}
\usepackage{ctable}
\usepackage{multirow}
\usepackage{balance}
\usepackage{booktabs}
\usepackage{hyphsubst}
\usepackage{graphicx}
\usepackage{amsmath}
\usepackage{url}
\usepackage{breakurl}
\usepackage{hhline}
\usepackage[caption = false]{subfig}

\newcounter{lizcounter}
\DeclareRobustCommand{\liz}[1]{\textbf{/* #1 (liz) */}\stepcounter{lizcounter}\typeout{LaTeX Warning: liz comment \thelizcounter: #1 (line \the\inputlineno)}}

\newcounter{findingscounter}

\newcommand{\para}[1]{\vspace{2mm}\noindent\textbf{#1}}

\setcopyright{rightsretained}

\begin{document}

\title{Overcoming the Imbalance Between Tag Recommendation Approaches and Real-World Folksonomy Structures with Cognitive-Inspired Algorithms}

\author{Dominik Kowald}
\affiliation{%
  \institution{Know-Center \& Graz University of Technology, Austria}
}
\email{dkowald@know-center.at}
\author{Elisabeth Lex}
\affiliation{%
  \institution{Graz University of Technology, Austria}
}
\email{elisabeth.lex@tugraz.at}


\begin{abstract}
Social tagging systems enable users to collaboratively annotate Web resources with freely chosen keywords (i.e., \textit{tags}). In order to assist users in this annotation process, tag recommendation algorithms have been proposed, which suggest a set of tags for a given user and a given resource \cite{P1}. Essentially, tag recommendation algorithms aim to help not only the individual to find appropriate tags \cite{jaschke2008tag} but also the collective to consolidate the shared tag vocabulary and thus, to reach semantic stability and implicit consensus \cite{wagner2014semantic}.

\para{Problem.}
While there is already a large body of research available with respect to tag recommendation algorithms, most existing approaches are designed in a purely data-driven way. Consequently, they rely on dense (i.e., broad) folksonomy structures, in which multiple posts per resource exist to have an appropriate amount of tag assignments available to train the algorithms. In order to ensure these dense data structures, many researchers use $p$-core pruning, which was, however, shown to unnaturally bias the tag recommendation evaluation process \cite{doerfel2013analysis}.

As indicated in Table \ref{tab:datasets}, the folksonomy structures of real-world social tagging systems are yet typically quite sparse (i.e., narrow), which is reflected by the average number of posts per resource (i.e, $|P| / |R|$). This degree of narrowness (i.e., sparsity) is smaller than two in four of the six systems. Thus, in these settings, current state-of-the-art tag recommendation approaches are not performing very well due to the lack of available training data \cite{P8}. This leads to an imbalance between the way how current approaches are designed and how the folksonomies of real-world social tagging systems are structured.

\para{Approach.}
To overcome this imbalance, we propose to build upon cognitive-inspired tag recommendation algorithms. We ground our approach on related research on the underlying cognitive mechanisms of social tagging, which has shown that the way users choose tags for annotating resources corresponds to processes and structures in human memory \cite{cress2013collective}. In this respect, a prominent example is the activation equation of the cognitive architecture ACT-R, which formalizes activation processes in human memory \cite{anderson2004integrated}. Specifically, the activation equation determines the probability that a specific memory unit (e.g., a word or tag) will be needed in a specific context. Thus, such an approach should enable us to efficiently model the tag vocabulary of a user in a cognitive-plausible way without the need for dense data structures associated with the resource to be tagged.

\begin{table}[t!]
	\small
  \setlength{\tabcolsep}{2.0pt}	
  \centering
    \begin{tabular}{l|ccccc|c}
    \specialrule{.2em}{.1em}{.1em}
											Dataset			& $|U|$		& $|R|$			& $|T|$							& $|Y|$     & $|P|$							& $|P| / |R|$	\\\hline 
											Flickr						&	9,590		&	856,755		&	125,119			&	3,328,590 &	856,755										& 1.000					\\
											CiteULike						& 18,474  & 811,175		& 273,883		& 3,446,650 & 900,794 								& 1.110					\\
											BibSonomy	  				& 10,179  & 683,478		& 201,254		& 2,986,396	& 772,108 								& 1.129					\\
											Delicious					& 15,980 	& 963,741		& 184,012			& 4,266,206 & 1,447,267 							& 1.501					\\
											LastFM							& 1,892		& 12,522		& 9,748			& 186,474		& 71,062									& 5.674					\\
											MovieLens					& 4,009		& 7,601			& 15,238			& 95,580    & 55,484									& 7.299					\\
		\specialrule{.2em}{.1em}{.1em}								
    \end{tabular}
    \caption{Statistics of publicly available datasets for Flickr, CiteULike, BibSonomy, Delicious, LastFM and MovieLens. Here, $|U|$ is the number of users, $|R|$ is the number of resources, $|T|$ is the number of tags, $|Y|$ is the number of tag assignments, $|P|$ is the number of posts and $|P| / |R|$ accounts for the degree of narrowness (i.e., sparsity).\vspace{-7mm}}	
  \label{tab:datasets}
\end{table}

Specifically, in \cite{P2,P3,P4,P5}, we have implemented a novel tag recommendation algorithm based on the activation equation of ACT-R. This algorithm ranks a user's tags based on three factors: (i) past usage frequency, (ii) past usage recency, and (iii) similarity with current semantic context cues. We have evaluated our approach using six real-world folksonomy datasets (i.e., Flickr, CiteULike, BibSonomy, Delicious, MovieLens and LastFM) and compared it to various state-of-the-art approaches using our \textit{TagRec} framework \cite{P6,P7}, which is freely-available on Github\footnote{\url{https://github.com/learning-layers/TagRec}}. This evaluation included popularity-based approaches (i.e, MP$_r$ and MP$_{u, r}$), Collaborative Filtering (CF), Latent Dirichlet Allocation (LDA), FolkRank (FR), an alternative recency-based approach (GIRPTM) \cite{zhang2012integrating} as well as our cognitive-inspired algorithm (ACT-R).

\para{Results.}
The results of this evaluation by means of the nDCG@10 metric are given in Table \ref{tab:results} \cite{P8,P9}. We see that the performance of the classic, data-driven tag recommendation approaches (i.e., MP$_r$, MP$_{u, r}$, CF, LDA, PITF and FR) strongly depends on the sparsity of the datasets. While their performance is quite weak for sparse (i.e., narrow) ones such as Flickr, CiteULike, BibSonomy and Delicious, they perform very well in dense (i.e., broad) settings such as LastFM and MovieLens. We observe a different behavior with respect to the time-dependent approach GIRPTM. Compared to the data-driven algorithms, this approach achieves the best results in the narrow folksonomy Flickr and the worst in the broad folksonomies LastFM and MovieLens. The overall best results are reached by our cognitive-inspired ACT-R approach.

\para{Discussion.}
Our results demonstrate that there indeed exists an imbalance between current state-of-the-art tag recommendation algorithms and the folksonomy structures of real-world social tagging systems. While these algorithms are designed for dense structures, most social tagging systems exhibit a sparse nature. To overcome this imbalance, we showed that cognitive-inspired algorithms, which model the tag vocabulary of a user in a cognitive-plausible way, can be helpful. Our present approach does this via implementing the activation equation of the cognitive architecture ACT-R, which determines the usefulness of units in human memory (e.g., tags).

Apart from that, in \cite{P10}, we have shown that ACT-R can be generalized for related use cases such as hashtag recommendations in Twitter. For future work, we plan to build upon this results in order to propose cognitive-inspired recommender systems (e.g., for resource recommendation) as an alternative to data-driven ones \cite{P11,P12,P13,lacic2015utilizing}. In this sense, our long-term goal is to design hybrid approaches, which combine the advantages of both worlds in order to adapt to the current setting (i.e., sparse vs. dense ones).

\para{Acknowledgments.} The authors would like to thank Stefanie Lindstaedt, Paul Seitlinger, Simone Kopeinik and Tobias Ley for supporting this research. This work is funded by the Know-Center Graz and the H2020 project AFEL (grant agreement: 687916).

\begin{table}[t!]
	\small
  \setlength{\tabcolsep}{2.1pt}	
  \centering
    \begin{tabular}{l|ccccccc|c}
    \specialrule{.2em}{.1em}{.1em}
					Dataset 			& MP$_r$		& MP$_{u, r}$				& CF				& LDA			& PITF				& FR			& GIRPTM 	& ACT-R			\\\hline 
					Flickr			  												  		                                              
												& -					& .569							& .666			& .280		& .535				& .561		& .686		& .711 			\\
					CiteULike		  										     		                                              
												& .063			& .392							& .359		  & .138	  & .294				& .392		& .422		& .438				\\											
					BibSonomy		  									        		                                              
												& .091			& .407							& .369			& .219		& .327				& .408		& .409		& .434					\\
					Delicious		  								          		                                              
												& .187			& .358							& .356			& .271		& .302				& .292		& .393		& .431					\\
					LastFM			  					                                                                  
												& .283			& .386							& .317			& .388		& .414				& .399		& .397		& .425							\\
					MovieLens		  					                                                                
												& .271			& .328							& .254			& .296		& .324				& .319		& .326		& .338							\\
		\specialrule{.2em}{.1em}{.1em}								
    \end{tabular}
    \caption{Evaluation results for various tag recommendation algorithms in real-world folksonomy structures by means of the nDCG@10 metric. Our cognitive-inspired approach (i.e., ACT-R) outperforms data-driven methods in all datasets.\vspace{-7mm}}	
  \label{tab:results}
\end{table}

\end{abstract}


\maketitle

\bibliographystyle{ACM-Reference-Format}

\end{document}